# Reading Convivial Solipsism: Towards a Logic of Perspectival Facts

*A clarification of common objections to non-absolute facts in Wigner-type scenarios*

Hervé Zwirn

*Centre Borelli (ENS Paris Saclay, France) & IHPST (CNRS, Université Paris 1)*
*herve.zwirn@gmail.com*

**Abstract:** Convivial Solipsism is easily misunderstood. Because it denies the absoluteness of observed events while preserving the universality of quantum mechanics, it invites objections that are often formulated in a classical, non-perspectival language. Some objections arise from importing assumptions that Convivial Solipsism rejects; others challenge whether those rejections yield an adequate account of observers, communication, and scientific practice. This article distinguishes these two kinds of criticism. In particular, it reconstructs a family of objections that arise in discussions of Convivial Solipsism, especially in connection with Wigner's friend and extended Wigner's friend scenarios. The objections are useful because they identify the conceptual discipline required by any interpretation that takes non-absolute facts seriously. The central claim is that observational facts must be indexed to perspectives. The mistake consists in removing these indices and then treating the resulting expressions as if they belonged to a single global factual domain. This leads to apparent contradictions and to false dilemmas concerning consciousness, communication, other minds, and intersubjectivity. When that is taken into account, it appears that Convivial Solipsism is a discipline of perspectival discourse. The paper also develops an analogy between two restrictions on global descriptions. Quantum contextuality prevents certain families of measurement statistics from being represented within a single noncontextual Kolmogorovian probability model. Likewise, Convivial Solipsism requires a perspectival logic governing the formation and combination of observational statements, while retaining classical reasoning for well-formed indexed propositions. A detailed formalization of the perspectival logic outlined here will be provided in a forthcoming paper.

## 1. Introduction: why misunderstanding ConSol is so easy

Convivial Solipsism (abbreviated ConSol) has been developed to solve the measurement problem and has been presented in a series of papers [1-11]. It belongs to the family of perspectival interpretations of quantum mechanics sometimes called Copenhagenish interpretations [12] alongside interpretations such as QBism [13-15], Relational Quantum Mechanics [16-19], Dieks's modal interpretation [20-22] or Brukner and Zeilinger's view [23-24]. Its central idea is not that the world is a private dream, nor that other observers do not exist, nor that physics collapses into psychology. Its central idea is that observed events, in quantum mechanics, are not absolute.

The reason why events are not absolute in ConSol comes from the way a measurement is defined: it is not a physical process but a perceptual one: a measurement is the selection at random in the observer's perception of one component of the entangled state vector attributed to the system and the measurement apparatus. The result of a measurement is then totally relative to the observer who performed the measurement[1]. Thus, events are always defined relative to the perspective (i.e. the observer's point of view) in which they are obtained, reported, or meaningfully expressed.

[1] To fully understand this paper, we recommend that readers have a basic understanding of Convivial Solipsism, which they can acquire by reading at least one of the following articles [3, 8, 9].

The fact that events are not absolute, while postulated in ConSol long ago, has recently found support through several no-go theorems that have been proven in the context of the so-called extended Wigner’s friend experiment [25]. These no-go theorems, while not identical, all point in the same direction: if we assume some reasonable hypotheses together with the assumption that events are absolute, then we are led to a contradiction. Therefore, at least one of the assumptions must be abandoned: either the absoluteness of observed events or one of the other hypotheses. The former is the choice made independently by ConSol and supported by these theorems[2].

This thesis is simple to state, but very difficult to respect in reasoning. The difficulty does not come only from the mathematics of quantum theory. It also comes from grammar. Ordinary language is built as if facts were, by default, shareable items in a common world. We naturally say "Alice saw spin up," "Bob read the string 0110," or "Wigner's friend obtained result A." Such sentences appear innocent. Yet, inside ConSol, they are incomplete. They tacitly omit the perspective relative to which the statement is being made.

The result is that many objections to ConSol arise from a subtle but decisive shift. The critic first speaks in the language of perspectival facts, but then translates those facts into a non-perspectival language. Once this has been done, contradictions, paradoxes, or metaphysical absurdities seem to follow. The correct diagnosis is that the apparent problem has been produced by an illicit change in logical type. The objections below are reconstructed arguments, some arising from discussions, rather than quotations attributed to particular authors.

This article aims to make explicit the rules that are often left implicit. It is written as a clarification of objections. These objections are not superficial. They concern the deepest issues raised by ConSol: what a perspective is, how another observer can appear in my perspective, what it means to deny the Tracking assumption, whether communication gives access to another observer's experience, whether the meta-level reintroduces an absolute viewpoint, and how ConSol differs from approaches that speak of the erasure or replacement of Wigner's friend's memory.

The guiding claim is this. ConSol should not be read as a metaphysical thesis saying that only my mind exists. It should be read as a discipline governing the formation and combination of observational statements. In that sense, ConSol is inseparable from a logic of perspectival facts that has been presented briefly in [11] and will be presented in detail in a forthcoming paper. The role of such a logic is not to replace ordinary logic everywhere. It is to prevent us from applying ordinary logical operations to expressions that are not well formed within a theory of non-absolute events.

Kolmogorovian probability remains applicable within a fixed measurement context but contextuality obstructs a global noncontextual representation of certain families of measurement statistics. The analogy proposed here concerns restrictions on combining perspective-indexed observational statements into a single observer-independent account which obstructs the use of classical logic because of the perspectival nature of events. Classical logic is not wrong; rather, it cannot be applied globally to observational statements that have not first been properly indexed to perspectives.

The paper proceeds as follows. Section 2 recalls the motivation supplied by Wigner's friend and extended Wigner's friend arguments. Section 3 explains the grammar of perspectival facts. Sections 4 to 12 examine the principal objections. Section 13 compares ConSol with Brukner's non-persistence of memory strategy. Section 14 presents the ontology. Section 15 points toward a formal logic of ConSol. Section 16 concludes by presenting ConSol as a discipline of perspectival discourse.

---

[2] We are not going to analyse these theorems and the way they should be understood inside ConSol here. This work has been done in [11].

## 2. Wigner, the friend, and the pressure against absolute facts

Wigner's friend scenario [26] is conceptually simple[3]. An observer seen by another observer can be in a superposed state like any other quantum system. A friend inside a laboratory measures a quantum system and, from their own point of view, obtains a definite result. Wigner, who remains outside the laboratory and treats the whole laboratory quantum mechanically, may assign an entangled state to the composite system consisting of the friend, the apparatus, and the measured system. The friend describes a definite outcome. Wigner describes a superposed or entangled state. The puzzle is not merely epistemological. The question is whether both descriptions can be taken as statements about one and the same observer-independent sequence of facts.

Extended Wigner's friend scenarios sharpen this tension. Brukner [24] formulated a no-go theorem for observer-independent facts, arguing that Wigner's and the friend's facts cannot always be jointly considered as local objective properties. Frauchiger and Renner [27] argued, in a related but different framework, that quantum theory faces a consistency problem when agents use the theory to reason about one another's reasoning. Bong and collaborators [28] later derived a strong no-go theorem involving local friendliness assumptions, including the idea that observed events are absolute. These results differ in formal structure, but they generate the same pressure: one cannot straightforwardly combine universality, locality-like assumptions, single outcomes, and the absoluteness of all observed events.

Well before these recent theorems, ConSol had chosen to abandon the absoluteness of observed events. It preserves the universality of quantum mechanics and accepts that an observer has a single definite experience, but it denies that all such experiences are facts in one global domain. The friend's result is not nothing. It is a result in the friend's perspective. Wigner's later description or report is not nothing. It is a result, or a record, in Wigner's perspective. The difficulty arises only if one insists on building a single observer-independent story containing both as unindexed facts.

This is why the debate is not only about physics. It is also about the rules of language. If the classical language of facts is retained unchanged, ConSol becomes almost impossible to understand. One will inevitably ask: What did the friend really see? Did Wigner really hear the friend's original result? Did the friend really have a perspective? Did the result really change? These questions already presuppose a single domain of factuality within which the answers should be placed.

ConSol's answer is not to say that the questions have no meaning at all. Its answer is to say that they must be reformulated. Instead of asking what the friend saw absolutely, one must ask what the friend saw in the friend's perspective, or what Wigner heard in Wigner's perspective, or what can be stated at the meta-level without dropping those indices. Once this is done, several standard objections lose their apparent force.

## 3. The grammar of perspectival facts

The minimal grammar needed for ConSol is the following. Let $O_i$ and $O_k$ be observers, and let j be a possible outcome. Let's stipulate that an expression such as $P(O_k^j)$ will be read informally as "observer $O_k$ saw outcome j." But in the strict language of ConSol this is not, by itself, a complete factual statement. It becomes a well-formed observational statement only when indexed to a perspective: $[O_i]\ P(O_k^j)$. This means: from the perspective of observer $O_i$, observer $O_k$ saw outcome j (which, when $i \neq k$, means that $O_i$ heard $O_k$ reporting having seen j). The special case $[O_i]\ P(O_i^j)$ expresses the observer's own outcome in that observer's own perspective. The point is not merely notational. It is constitutive.

[3] For the reasoning we are going to present, it is generally assumed that quantum mechanics is universal, which means that it applies to all scales including the macroscopic one. Every physical process is unitary and there is no collapse.

Without the perspective index, the statement behaves as if it belonged to a global factual domain. ConSol denies that such a domain is available for observed quantum events.

This grammar allows an observer to speak about what appears in their perspective. Thus Wigner can say, within Wigner's perspective, that the friend answered "up." The friend can say, within their own perspective that they saw "down." What ConSol blocks is the inference that both statements can be immediately inserted into a single narrative of what really happened, independently of any perspective.

The same point applies to negation. A statement such as [Wigner] not P(friend$^A$) says that, from Wigner's perspective, it is not the case that the friend reports having seen A. This negation concerns the report as it appears in Wigner's perspective. The friend may report another result, remain silent, or not yet have been questioned; the negation alone does not determine which of these possibilities obtains. That does not tell us anything about what the friend experienced in their own perspective. Negation must be handled within the expressive resources of the perspective in which it occurs.

This discipline resembles indexical logic [29, 30, 31] in one respect. Just as "I am here now" cannot be evaluated without a context of utterance, "the friend saw A" cannot be evaluated without a perspective. But ConSol is more radical than ordinary indexicality. In standard indexical semantics, once the context is fixed, the indexical expression often yields an ordinary proposition that can be placed in a common world. In ConSol, fixing the perspective does not license the removal of the perspective. The fact remains perspectival.

A meta-level is necessary for expressing some subtle points and can be introduced, but it must be carefully understood. The meta-level allows the Modeller to juxtapose indexed results obtained in the perspectives of the observers involved in the context under consideration. Such a juxtaposition is a family of perspectival records, not a global perspective, and it does not turn indexed facts into absolute facts. It is not accessible to any observer (see Section 9 for what it does and does not commit us to). The Modeller represents such a juxtaposition at the theoretical level; this representation does not imply that any observer can access the experiences indexed to other perspectives.

The meta-level allows the Modeller to discuss and compare indexed propositions. The Modeller is not an observer, so sentences stated by the Modeller will be preceded by the prefix [[Meta]] to distinguish them from sentences stated by an observer, which are preceded by the prefix of a perspective such as [$O_i$].

It is important to understand that the Modeller, using the meta-level, speaks about indexed sentences while observers, using the object language, speak of events.

No observer can make a statement about the results obtained by another observer in that observer's own perspective. Sentences mixing sentences indexed in different perspectives like {[$O_i$] P($O_i^k$) and [$O_j$] P($O_j^l$)} are not allowed. But it is possible to write something like [[Meta]]{[friend] P(friend$^A$) and [Wigner] P(friend$^B$)}. This means that in the friend's perspective, the friend saw the result A and in Wigner's perspective, the friend is assigned result B. The meta-level allows the Modeller to state sentences mixing sentences indexed in different perspectives, which is not allowed inside one given perspective.

Notice that the meta-level does not authorize the inference from [[Meta]]{[friend] P(friend$^A$)} to [[Meta]] P(friend$^A$) as an unindexed absolute fact. It does not re-establish a God's-eye view accessible to any real observer.

This language and the associated perspectival logic will be defined more precisely in a forthcoming paper.

The central assumption, Tracking, can be stated only at the meta-level. It is defined as:

Tracking: [[Meta]] {if $[O_i]$ $P(O_k^j)$ then $[O_k]$ $P(O_k^j)$}.

It is possible to state Tracking more precisely using modal logic (as we explain in [11]):

[[Meta]] {$\Box$ ($[O_i]$ $P(O_k^j)$ => $[O_k]$ $P(O_k^j)$)}.

In words: if, from $O_i$'s perspective, $O_k$ reports outcome j, then, in $O_k$'s own perspective, $O_k$ has outcome j. This assumption is extremely natural. That means that when you hear somebody telling you that they obtained a given result, this person has really obtained this result in their own perspective. It is precisely what ConSol rejects as a general principle. The violation of Tracking means that the content attributed to $O_k$ within $O_i$'s perspective need not coincide with $O_k$'s own first-person content. A possible violation of Tracking, when both indexed statements attribute definite outcomes, can be expressed as follows:

[[Meta]] {$\Diamond$ ($[O_i]$ $P(O_k^j) \wedge$ $[O_k]$ $P(O_k^l)$)} with $i \neq k$ and $j \neq l$

## 4. Objection number 0: Thinking that ConSol is a solipsistic interpretation

This is labelled objection number 0 because, unlike the following objections, which are assigned a proper number, it is superficial even if persistent. It is true that the name “Convivial Solipsism”, which is an oxymoron, could suggest at first sight that this interpretation is solipsistic. This, however, is not the case. Solipsism is the theory according to which only one person exists - myself - and everything else comes from me. ConSol allows all the other observers to exist independently of me. On the minimal reading of Section 14, this independence is itself a structure articulated within experience; on the realist reading it is ontological. One of the main questions studied in ConSol is precisely the communication between observers! The reason why the name of the interpretation contains solipsism is only related to the fact that each observer lives in their own perspective (or bubble) without any possibility of having direct access to other perspectives. The adjective “convivial” refers to the consistency of the measurement reports perceived within an observer’s perspective, under ideal conditions (see section 8).

## 5. First objection: treating a perspective as an object possessed by an observer

One common objection begins as follows: if ConSol speaks of several perspectives, then one should be able to ask whether the observer appearing in my perspective has a perspective. If the observer has one, the critic says, then Tracking is not violated. If the observer does not have one, then ConSol appears to deny other minds.

This objection treats a perspective as if it were an object that an observer possessed, rather like a car, a brain, or a memory register. That is not how the term functions in ConSol. A perspective is not an additional entity inside the world. It is not something that I inspect in another observer and then attribute to them. It is the logical index relative to which observational statements become meaningful.

To associate a perspective with an observer is to say that statements expressing that observer's own experience are to be evaluated relative to that observer's first-person standpoint. It is not to say that the observer carries around a metaphysical object called a perspective. The perspective index marks the standpoint relative to which statements are evaluated and the corresponding experiential history of the observer’s successive outcomes. It is not an item among the objects described by the statement.

Consider the statement: "Bob feels pain." In ordinary language, this statement looks complete. In ConSol, if it is an observational or experiential statement, it must be indexed. There is a difference between [Bob] “Bob feels pain” and [Alice] “Bob appears to be in pain”. The first concerns Bob’s first-person experience. The second concerns what appears in Alice’s perspective. The question is not whether Alice can see Bob's perspective as an object. The question is which indexed statement is being made.

The same applies to Wigner's friend. It is misleading to ask whether the friend appearing in Wigner's perspective has, inside Wigner's perspective, a perspective of their own. Wigner can describe a person, a body, a memory register, or a report appearing in Wigner's perspective. But the friend's own perspective is not another object inside Wigner's perspective. It is the index under which statements of the friend's experience are formed.

This clarification dissolves a false dilemma. ConSol does not need to say either that the other observer has no perspective or that the other observer's perspective is directly accessible to me. It says that we cannot speak within our own perspective of the other observer's own experience. If the other observer's own experience is mentioned, then it is at the meta-level only. When I speak of what the other observer appears to say or do within my experience, I must use my perspective index. Confusing these two levels is precisely the error. One confusing issue is that when we discuss the interpretation, we often switch between statements that we utter as simple observers and statements that we utter as people describing the interpretation. In the latter case, we speak as the Modeller.

A stronger objection does not treat a perspective as an object. It asks what relates the interlocutor appearing in my experience to the subject whose first-person experience is discussed at the meta-level. ConSol identifies the interlocutor appearing within an observer’s perspective with the observer whose first-person experience is discussed at the meta-level. This identification is part of the theoretical description, made at the meta-level: the same observer label occurs under different perspective indices. However, identifying the observer does not entail identifying the experiential contents attributed to that observer under those different indices. The identification of the subject and the correspondence of content are thus two distinct propositions, both formulated at the meta-level. The first does not entail the second.

## 6. Second objection: inferring identity of experiential content from identity of the observer

A second objection is closely related to the first. It consists in assuming that the observer who appears in my perspective must be identical, in all relevant observational content, with the observer in their own perspective. This assumption seems obvious in everyday life. If Bob stands in front of me and says "I saw A," I normally take Bob's utterance to reveal Bob's own experience. In ordinary circumstances that assumption is indispensable. This is precisely the Tracking assumption. But in the context of Wigner-type quantum scenarios it is exactly what must be questioned.

ConSol distinguishes between two roles. There is the observer-token as it appears in my perspective: the Bob I see, the voice I hear, the message I read, the record I obtain. And there is Bob's own first-person perspective: the standpoint from which Bob's own experiential statement is meaningful. The ordinary assumption is that these two are automatically connected by a reliable identity of content. ConSol denies that this assumption (Tracking) is generally legitimate.

This denial does not fragment persons into a collection of zombies. It refuses an inference concerning observational content. The Bob who appears in my perspective may be described as “saying A”. But from that statement alone one cannot infer that [Bob] “Bob saw A”. One can infer only [Me] “Bob reports A”. The additional step from the report-as-it-appears-to-me to Bob's own experience is precisely Tracking.

The critic may object that this is unintelligible. If Bob is conscious, surely the report I hear from Bob must express Bob's consciousness. But this objection simply restates the classical picture. It assumes that communication functions as a transparent channel from one perspective to another. In a theory of non-absolute facts, that assumption must be earned, not presupposed.

The issue is not whether Bob exists or whether Bob is conscious. The issue is whether the content attributed to Bob within my perspective is necessarily identical to the content of Bob's own experience. To say that this identity is not guaranteed is not to deny Bob's first-person life. It is to deny that my access to Bob is trans-perspectivally transparent.

A useful analogy comes from indexical contexts, though it remains imperfect. If Alice says "I am here," the word "here" refers to Alice's location in Alice's utterance. If Bob reports "Alice said 'I am here'", the indexical content must be handled carefully. We cannot simply move the content into Bob's context without adjusting the index. ConSol requires a much deeper caution: the index is not just linguistic. It marks the domain in which an observational fact is formed.

## 7. Third objection: violation of Tracking as denial of other minds

The most serious objection says: if Tracking is violated, then the observer whom I see reporting an outcome cannot have a genuine perspective. If they had a perspective, then what they report would be what they experience, and Tracking would hold. Therefore, the objection concludes, ConSol either denies the perspectives of others or cannot coherently claim that Tracking is violated.

This objection is powerful only because it builds Tracking into the very notion of having a perspective. But that is exactly what is at issue. The existence of another perspective does not imply that my attribution of content to that perspective is correct. Nor does it imply that a report appearing in my perspective transparently expresses the other perspective's content.

Let us write the situation explicitly. Suppose that, in my perspective, I saw a certain result r and hear Bob say: "I saw the result r." The well-formed statement available to me is:

[Me] "I hear Bob reporting that he saw the same result r as I did".

The critic wants to infer:

[Bob] "Bob saw the result r".

That inference is not a mere consequence of Bob's being conscious. It is a rule, rejected by ConSol, connecting a report in my perspective with an experience in Bob's perspective.

A violation of Tracking is therefore not the claim that Bob lacks a perspective. It is the claim that the content of Bob's report as it appears in my perspective may differ from the content of Bob's own first-person experience. Bob may have a perfectly genuine perspective. What is denied is my right to treat the Bob-report in my perspective as a window onto that perspective.

The critic might reply that this makes communication useless. But that reply exaggerates the point. Communication remains fully functional within a given perspective. I hear, read, respond, compare, write, and construct stable records[4]. I can even experience complete agreement with others. What ConSol denies is not experienced agreement but absolute trans-perspectival identity.

Thus the violation of Tracking does not create mindless interlocutors. It creates a restriction on what may be inferred from the appearance of an interlocutor. It is an epistemic and logical restriction, not a denial of consciousness.

[4] It is worth noting here, as emphasized in section 8, that the hanging-on mechanism requires that, after a measurement, the observer's perception remains anchored to the branch that was selected for any subsequent measurement. This ensures the stability of perceptions.

## 8. Fourth objection: communication as direct access to another perspective and the common bit string as an absolute anchor

The natural picture of communication is transfer. Alice has a fact in her mind. She expresses it. Bob receives it. The same content is now available to Bob. This picture works well enough in ordinary life. It is also embedded in many discussions of Wigner's friend. Wigner asks the friend: "What did you see?" The friend answers. Wigner is then assumed to know the friend's result.

In ConSol, this picture is not generally valid. Communication is itself an interaction. If Bob interrogates Alice, then he makes a measurement on Alice. What he hears Alice saying is therefore a result that belongs to Bob's perspective. If Wigner asks the friend and hears "A," the well-formed statement is not [Friend] "Friend saw A". It is [Wigner] "Friend answered A". The latter is a fact in Wigner's perspective. The former is a fact in the friend's perspective. The inference from one to the other is Tracking.

The reason why Tracking is not guaranteed is easy to see once we recall what a measurement is in ConSol. It is the random selection, in the observer's perception, of one of the possible results present in the superposed state vector of the system, but the state vector is not changed. There is no collapse. So there is no reason why this random selection should be the same for two different observers because neither the formalism of quantum mechanics nor ConSol imposes a cross-perspectival rule requiring these selections to coincide. This does not mean that communication is illusory. It means that communication does not perform a miracle. It does not transport a fact from one perspective into another while preserving its identity independently of all perspectives. It produces a new event in the receiving perspective: an utterance heard, a message read, a pointer seen, a memory updated.

The point can be illustrated by the case of a bit string. Suppose Alice produces a random string in her perspective, writes it down and sends it to Bob. Bob later reads the string in his perspective. In a classical setting we assume that the written string and the read string are the same object. In a Wigner-type quantum setting, ConSol denies that this sameness can be taken for granted at the level of observed facts. The statement "the same string" itself must be indexed.

The critic may insist: surely we can define the string explicitly, for instance as S = "1100001110111101010111". Either Bob saw S or he did not. If both participants can see the same written sequence, then there must be a common fact anchoring the debate. A bit string seems to be an ideal candidate for intersubjective reference. It can be copied, printed, compared, transmitted, and checked. This seems to defeat the idea that facts are perspectival.

But this does not solve the problem. "This string" is not a magical device that bypasses perspective. The symbols that Alice writes, the symbols that Bob reads, and the symbols that appear in the subsequent conversation are all events in perspectives. The apparent identity of the inscription does not by itself establish a trans-perspectival identity of the observed content.

The unindexed statement "Bob saw S" is not a well-formed observational statement. The well-formed statements are [Alice] "Alice saw $S_A$" and [Bob] "Alice appears to have written $S_B$", where $S_A$ and $S_B$ may be represented within different perspectives. If one asserts that $S_A = S_B$ as a trans-perspectival fact, one has added a Tracking-like identification.

Communication, in ConSol, therefore has a double status. Within a perspective, it is real and stable. Across perspectives, it does not automatically ground identity of content. The classical assumption that it does so is precisely what ConSol declines to impose as a general principle.

This is why the slogan "Wigner asks the friend" is dangerous. It compresses at least two different statements. First, in Wigner's perspective, Wigner experiences asking the friend and hearing an answer.

Second, in the friend's perspective, the friend may have their own experience of the measurement and of the conversation. ConSol does not deny either statement. It denies that they can be fused without further assumption.

Nevertheless, it should be added that inside their own perspective, an observer who interrogates another observer hears an answer consistent with their own result. This is the convivial aspect of the interpretation. That follows from two features: the hanging-on mechanism[5] which guarantees that the observer's perception remains anchored to the selected branch for the following measurements and the fact that asking another observer a question is measuring them. This consistency is a property of what appears in the interrogating observer's perspective; it does not entail that the answer expresses the interrogated observer's own first-person result, which is precisely what Tracking would assert.

## 9. Fifth objection: does the meta-level reintroduce an absolute viewpoint?

This is probably the most serious structural objection. It runs as follows. ConSol states its own rules, including Tracking and its violation, in a [[Meta]] language that compares results indexed to different perspectives. Such comparison presupposes that the results of all perspectives are jointly given, which is exactly the global factual domain that ConSol denies and that the no-go theorems target. Either the meta-level is idle, or it smuggles in the absoluteness it was meant to remove. The objection deserves a precise answer, in three steps.

Let's first say that the metalanguage is the language of the Modeller who is not an observer but a theoretical commentator.

(i) The role of the meta-level. The meta-level is a metalanguage in the ordinary logical sense: a language for talking about indexed sentences, their well-formedness, and the inferences between them. It allows us to say that a given indexed statement belongs to a given perspective, that two indexed statements belong to different perspectives, that a combination of them is not permitted within any single perspective, and that a given inference requires Tracking. Its governing constraint is that it never removes an index. The meta-level may contain:

[[Meta]] [Alice] "Alice saw A",

but not, as an observational fact:

[[Meta]] "Alice saw A".

Indexed facts are true or false as indexed facts. The truth of an indexed fact does not entail the truth of an unindexed one, and the meta-level contains no rule that licenses this passage.

(ii) What a juxtaposition of indexed results is, and is not. The objection treats the Modeller's comparison as an assignment of one outcome to each observer within a single probability space. This is not what ConSol needs, and the difference matters. What the no-go theorems exclude is the conjunction of several assumptions, among them that observed events are absolute: each has a single outcome that holds irrespective of any perspective, and these outcomes are jointly governed by one distribution satisfying the other assumptions of the theorem. A juxtaposition of perspectival records has none of these properties. It is a family of local items, each of which is meaningful within its own perspective. Nothing in ConSol requires that this family be the marginal of a joint distribution over all perspectives, or that it be the restriction of a single assignment of facts. The only probabilistic constraints that ConSol imposes are those that quantum theory fixes within each perspective. The comparison of indexed records does not, by itself, license an inference from a report in one perspective to an experience in another; such an inference requires an explicitly postulated bridge. The meta-level therefore describes a collection of local sections, not a global one.

[5] The argument has been given explicitly in several previous papers, see for example [3].

(iii) What ConSol commits to, and what remains open. The Modeller is a theoretical role, not an observer. The comparisons it makes are not accessible from any perspective, and they do not provide a vantage point from which “what really happened” could be read off. The justification for the meta-level is functional: without it, one could neither state Tracking nor reconstruct the assumptions of the extended Wigner’s friend arguments in order to say which one is rejected. A test follows from this. Any argument that moves from a meta-level statement to an unindexed observational statement without an explicitly stated bridge has left ConSol, and the burden of justification lies with whoever makes the move.

A further point should be made explicit. The theory does not need the assumption that a *complete* table of indexed results exists for all observers at once. What it needs is that, for any finite collection of perspectives under consideration, the Modeller can compare their indexed statements. The relevant distinction is not between finite and infinite collections, but between comparing indexed statements and identifying their contents across perspectives. This weaker requirement suffices for every use made of the meta-level in this paper. Whether a complete table should be postulated at all is a question on which ConSol can remain agnostic, and the choice does not affect the empirical content of the theory. The two readings of the ontology (Section 14) do differ on how to interpret such comparisons. On the minimal reading they are bookkeeping devices of the theory. On the realist reading they describe relations between empirical realities that are each relative to an observer.

The meta-level is therefore not an exception to perspectivalism. It is the place where the rules of perspectival discourse are formulated, and whoever rejects it must explain how Tracking and its violation can otherwise be stated.

## 10. Sixth objection: ConSol as a rejection of non-contradiction

Another objection is that ConSol solves paradoxes by weakening logic. If Alice sees A and Bob sees B, and if A and B are incompatible, then perhaps ConSol allows contradictions. This would make it a paraconsistent interpretation, or worse, a refusal of rational constraint.

This is not correct. ConSol does not reject non-contradiction. It rejects ill-formed combinations of propositions. The difference is crucial. A contradiction requires two propositions belonging to the same logical space. But if [Alice] “Alice saw A” and [Bob] “Bob saw B” are indexed to different perspectives, they cannot simply be treated as A and B in one unindexed domain.

Within a given perspective, ordinary consistency constraints remain and a proposition and its negation cannot both be consistently asserted. If [Alice] “Alice saw A” is true, then [Alice] “not (Alice saw A)” cannot also be asserted in the same sense and in the same respect. At the meta-level too, contradictory claims about the same indexed formula are not both acceptable. ConSol is not a free-for-all.

What ConSol blocks is the construction of contradictions by de-indexing. Suppose one has [Alice] “Alice saw A” and [Bob] “Bob saw B”. If one converts these into "Alice saw A" and "Bob saw B" as absolute facts and then adds further assumptions connecting the two (for example that these results are for the same observation), a contradiction may emerge. But the contradiction belongs to the illicit reconstruction, not to ConSol.

This is similar in spirit to errors involving context-sensitive terms. If Alice says "I am tired" and Bob says "I am not tired," there is no contradiction unless the indexical term "I" is mishandled. The two utterances concern different persons. ConSol generalizes this caution in a more radical way. Observational outcome statements are not complete without a perspective.

The lesson is not that logic fails. It is that logic needs well-formed inputs. Classical non-contradiction applies once the relevant propositions have been correctly formed. It does not authorize us to erase the very indices that make them meaningful.

This point is important for the interpretation of extended Wigner's friend no-go theorems. The contradiction-like pressure in such arguments often arises from combining statements made by different agents under assumptions that effectively turn those statements into elements of one global factual domain. ConSol's response is to deny the legitimacy of that global domain for observed events. The logical rule is preserved, but the domain of application is refined.

## 11. Seventh objection: ConSol as the denial of intersubjectivity

Because ConSol denies absolute observed facts, it is sometimes accused of destroying intersubjectivity. Science, it is said, depends on public facts. If every fact is trapped inside a perspective, then no common inquiry is possible. This is a strong objection stated by Adlam [32] against every perspectival interpretation.

As discussed in detail in [11], this accusation rests on a false alternative. Either facts are absolute and science is possible, or facts are perspectival and science collapses. ConSol rejects this alternative. Scientific objectivity need not be grounded in a God's-eye collection of facts. It may be grounded in the coherence and stability of perspectival structures as experienced within perspectives[6].

From within my perspective, I encounter other observers, reports, instruments, papers, conversations, experiments, and agreements. These are not illusory. They form the empirical world in which scientific practice occurs. I can repeat an experiment, exchange records, calculate frequencies, correct errors, and develop theories. All of this remains intact within the perspective.

What ConSol denies is the additional metaphysical claim that these practices must be underwritten by an observer-independent set of all observed facts. It is one thing to experience stable agreement. It is another thing to claim that the experienced agreement reveals an absolute identity of facts across perspectives.

This distinction is delicate because ordinary scientific realism naturally seeks a stronger notion of objectivity. It wants the facts to be the same for everyone, independently of any standpoint. ConSol argues that quantum mechanics may force us to weaken this demand. The weakening is not arbitrary. It is motivated by the same formal tensions that appear in extended Wigner-type scenarios.

Intersubjectivity in ConSol is therefore not abolished but reinterpreted. It is not the direct sharing of an absolute fact. It is the internal coherence of a perspective in which other observers appear as reliable interlocutors and experimental practice remains stable. This may seem philosophically costly. But it may be less costly than forcing all observations into a single factual domain that quantum mechanics itself appears to undermine.

One may compare this with other limits discovered in science. The fact that there is no absolute simultaneity in relativity does not destroy temporal order altogether. It forces us to index time-related statements to frames. Similarly, the non-absoluteness of quantum events does not destroy empirical discourse. It forces us to index observational statements to perspectives.

[6] See section 8 in this paper and [11] for a more detailed analysis of this point.

## 12. Eighth objection: ConSol is either a version of the old interpretation of von Neumann, Wigner, and London and Bauer, or a version of the single-mind interpretation of Albert and Loewer

The role of consciousness inside ConSol is totally different from the role that von Neumann, Wigner, and London and Bauer [33] assumed it to play. At least on some readings of their view, their interpretation is dualist and consciousness is assumed to be a nonphysical entity. Nevertheless, consciousness is supposed to be responsible for the physical collapse of the wave function[7]. Nothing similar appears in ConSol.

ConSol is also different from Albert and Loewer's single-mind interpretation [34, 35]. In this interpretation, which is a version of the Everett interpretation, the universal wave function always evolves unitarily and is the same for all observers. All terms in the superposition exist physically, but each observer has a unique mind that follows a single branch. Furthermore, the mind is a non-physical entity added to the physical description. There may be instances where Bob's body is present, but Bob's mind is not. This raises the problem of "mindless hulks": bodies without minds. This difficulty motivated Albert and Loewer's development of the Many Minds interpretation. But the most important difference is the existence of absolute facts: the fact that Alice's mind followed that particular branch is an absolute fact. Therefore, the fact that Alice saw "up" is an absolute fact that holds true for everyone. In the Single-Mind version, mental trajectories are ontological elements of the model, whereas ConSol refuses to synthesize the perceived results into a single, absolute narrative. The single-mind interpretation adds a non-physical mind to an Everettian ontology to explain why a given observer experiences only one outcome; ConSol maintains that the outcome itself has meaning only within a particular perspective.

## 13. Brukner and the non-persistence of memory: a different strategy

The comparison with Brukner's approach is interesting. Brukner's no-go theorem for observer-independent facts already points toward the idea that facts may not be common to Wigner and the friend [24]. Later work by Guérin, Baumann, Del Santo, and Brukner [36] formulates a no-go theorem for the persistent reality of Wigner's friend's perception, arguing that the perceptions the friend has of her own measurement outcomes at different times cannot always be treated as sharing the same reality under natural quantum assumptions. Baumann and Brukner further discuss Wigner's friend's memory and the no-signaling principle, emphasizing that Wigner's measurement can change the internal record of the friend's initial result [37].

This suggests a strategy different from ConSol. Suppose the friend initially obtains A. Later Wigner performs an operation on the whole laboratory and obtains, or elicits, B. In Brukner's line of thought, the friend's initial record may fail to persist. The later interaction may erase, modify, or replace the friend's memory of the initial result. At the end of the process, Wigner and the friend may share a new result, while the previous result no longer survives as a stable memory.

ConSol does not need to tell this story. This concerns the explanation of cross-perspectival discrepancies; it does not imply that ConSol forbids physical changes to memory registers caused by Wigner's intervention. In ConSol, the friend's initial result remains meaningful as [Friend] "Friend saw A". Wigner's later experience remains meaningful as [Wigner] "Friend reports B". The problem is not solved by erasing the first result. It is solved by refusing to put both results into a single absolute temporal narrative.

[7] In fact, von Neumann was never explicit about that and London and Bauer's position appears much more subtle than that when one analyses what they have written. But it is the usual (and naïve) way to present their interpretation and this is the way it is understood by a vast majority of physicists who, of course, reject it at first sight.

A more detailed analysis of the differences between ConSol and Brukner's position will be proposed in a forthcoming paper.

### 14. What is the ontology?

This section is necessarily much more metaphysical and risky. The goal is to give the most precise possible description of the ontology of ConSol. Such an effort is rarely undertaken explicitly in discussions of quantum interpretations. QBism, for example, takes agents' experiences and interactions with external systems as starting points without attempting to provide a detailed ontology of the underlying processes [6].

ConSol does not impose a cumbersome ontology. Its minimal core is sufficient to account for the non-absolute nature of events. However, a more realist interpretation is available for those who wish to maintain a more vivid and explanatory understanding of measurement. These two readings, although different in their metaphysical meaning, yield the same empirical predictions within the framework of ConSol.

The minimal ontology is limited to observers, each embedded in their relative external world (which is left undescribed) and able to interact with this external world. These physical interactions give rise to perceptions of results, and these perceptions are precisely what ConSol calls measurements. To each observer is attached a global wave function summarizing the possible results and the probability of their occurrence. As I explain in [11]:

> *Three key notions are perspective, observer, and meta-level description. A perspective P is defined as a temporally ordered sequence of recorded outcomes, corresponding to the successive perceptual experiences of a single observer. Each perspective provides a coherent account of measurement results constituting the phenomenal reality of the observer, and all statements about outcomes are meaningful only relative to such a perspective. An observer is a stable structure with the ability to engage in interactions that produce further outcomes attached to a perspective, characterized by memory and consistency. A meta-level description is a theoretical construct used to compare or relate different perspectives. It does not correspond to the viewpoint of any physical observer and does not provide access to a global state of affairs.*

This minimal ontology does not postulate anything about the external world of the observers. It does not even postulate anything about the existence of the observer beyond the very primitive concept of experience. This version is close to the most radical version of QBism as described by Bitbol in the pure phenomenological tradition [38]:

> *[…] neither the agent nor physical systems are regarded as things that truly exist outside of us. For the agent equipped with measuring instruments, just as much as physical systems, represents nothing more than internal aspects of experience. This does not prevent something like the world—or, more vaguely, the explored environment—from transcending the agent's personal experience. But this transcendence is itself encoded within the immanence of present experience.*

In this minimal reading, observers and their relative external worlds are understood as structures articulated within experience, rather than as independently postulated substances. This version is radical in the sense that it pushes the phenomenological approach to its logical conclusion by practicing the *épochè*, which consists in suspending all judgement concerning the existence of a reality (the empirical reality) that transcends our perceptions beyond the phenomenological reconstruction of a reality that is limited to the minimum field directly resulting from our perceptions (the phenomenal reality).

The second reading takes the global wave function to describe something called "empirical reality", which is the part of the inaccessible external world that is closest to the observer (in a way, the tip of the iceberg) with which the observer can interact through the experiments they perform. Thus, the empirical reality, which is supposed to be real in a certain sense, is described by the entangled state vector and evolves linearly and deterministically according to the Schrödinger equation. This evolution remains unitary and involves no physical collapse. But given the perceptual limits of the observer, it is only possible for them to perceive certain components of this empirical reality, which is not accessible to them in all its superposed richness. A measurement therefore consists of taking a look at this empirical reality through an experiment and perceiving a section of this empirical reality which will be considered as the observer's phenomenal reality, whereas the perceptual selection introduces no additional physical change beyond the unitary evolution associated with the experiment.

This second presentation, which is close to d'Espagnat's conception of a veiled reality [39], will suit those who wish to rely on an image closer to a certain type of realism. It also provides a more explicit solution to the measurement problem since a measurement is no longer a physical process governed by unitary evolution but is a mental process constrained by the filters of our perception.

On the minimalist reading, the wave function is only a structured catalogue of possible experiences and their respective probabilities. On the realist reading, it represents an empirical reality, not fully accessible but real in a strong sense, although relative to the observer and never reducible to a shared Everettian universal wave function.

Since every possible test ultimately results in a phenomenal event for some observer, and since both readings agree on such events, the minimalist and the realist versions of ConSol are empirically indistinguishable. The choice is therefore not a scientific choice in the strict empirical sense. It is a metaphysical choice underdetermined by the evidence and by quantum mechanics. The minimalist version is more parsimonious, while the realist version offers a more explicit explanatory picture, especially regarding why measurement appears as the selection of a single classical outcome from a superposed empirical structure. Depending on their philosophical assumptions, readers are therefore free to choose either of these two versions.

## 15. Toward a formal logic of ConSol

The preceding discussion points toward a formal project. If ConSol is to be more than an interpretive slogan, it needs a logic of perspectival facts. Such a logic defines which expressions are well formed, how perspectival operators behave, what can be stated at the meta-level, and which inference rules require additional assumptions such as Tracking.

The first rule is formation. Observational predicates require a perspective index. $P(O_k^j)$ by itself is not well formed. $[O_i]\ P(O_k^j)$ is well formed. This single rule prevents many confusions.

The second rule concerns communication. From [$O_i$] "$O_k$ reports j" one cannot infer [$O_k$] "$O_k$ experienced j" unless Tracking is assumed. Communication inside a perspective is not trans-perspectival access.

The third rule concerns meta-description. The meta-level does not speak of events; it speaks of indexed propositions. It may combine and discuss them, but it may not eliminate their indices. [[Meta]][$O_i$] "P" is correct. [[Meta]] "P" is not correct when P is an unindexed observational fact.

The fourth rule concerns non-contradiction. Classical consistency applies within each perspective and at the meta-level for well-formed formulas. But contradictions can be obtained by treating formulas from different perspectives as if they were unindexed formulas in one domain.

The fifth rule concerns identity of outcomes. A symbol, string, or record appearing in two perspectives cannot be assumed identical as an observed fact merely because the same external notation is used. Identity across perspectives is itself a substantive claim requiring justification.

These rules are modest, but they change the structure of the debate. They force critics to state explicitly when they rely on a trans-perspectival bridge. The most important such bridge is Tracking. Once Tracking is made explicit, the disagreement becomes clear. The critic may defend Tracking. ConSol rejects it as a general principle.

A more fully developed formal system will be given in a forthcoming paper. But even this minimal grammar is sufficient to block the most common objections.

The aim is not to create an exotic logic for its own sake. The aim is to prevent a classical syntax from distorting a non-classical interpretation. If the physics suggests that events are not absolute, then the language of facts must be adjusted accordingly.

## 16. Conclusion: Perspectivalism as a new form of Contextuality

The aim of this paper was to separate two kinds of objection to ConSol: those that derive from de-indexing, and those that challenge whether the rejection of absolute observed events yields an adequate account of observers, communication, and science. The first kind dissolves once indices are kept. The second kind concerns the adequacy of the account itself and is at least partly addressed by the present clarification.

What ConSol says. Observational statements are well formed only relative to a perspective. From the statement that an observer reports an outcome in my perspective, no statement about that observer's own experience follows without an additional bridge, which I have called Tracking. ConSol rejects Tracking as a general principle, but it does not deny that other observers exist or have experience, does not accept contradictions, and does not hold that communication fails. Within a perspective, communication, records, and experimental practice are stable, and ordinary logic applies to well-formed indexed formulas. The meta-level is a metalanguage that compares indexed statements without removing their indices. It supplies no global perspective and does not require a joint assignment of outcomes across perspectives.

The rules proposed in Section 15 make the logical structure of the disagreement explicit. A critic who finds ConSol unacceptable may defend Tracking, or may argue that a global domain of observed facts is indispensable, and in both cases the debate bears on a stated assumption instead of on an implicit one. This clarifies the position, but it does not establish its truth.

These objections nevertheless help define the positive theory. ConSol requires not only a metaphysics but also a method. The method is: never remove an index without justification; never infer another perspective's content from a report in one's own perspective without Tracking; never treat communication as transparent access; never use the meta-level to restore absolute facts; never manufacture contradictions from ill-formed de-indexed propositions.

Perspectives are indices of discourse and domains of well-formed observational statements. To speak from a perspective is to speak under a rule: one may describe what appears in that perspective, but one may not thereby claim direct access to another perspective's first-person content.

The meta-level does not undermine this view. It is not a God's-eye standpoint. It is a formal device allowing us to discuss indexed propositions without removing their indices. Nor does ConSol reject non-contradiction. It preserves consistency for well-formed formulas and blocks contradictions produced by de-indexing.

Finally, it is worth noting that quantum contextuality showed that the classical probability calculus cannot be applied globally to quantum events because such events do not form a single Boolean algebra. The probability calculus appropriate to quantum events is not a single global Kolmogorovian model, but the quantum calculus of probabilities. The non-absoluteness of observed events suggests an analogous limitation at the level of factual discourse: classical logic cannot be applied globally to observational facts if such facts do not belong to a single observer-independent domain. There is a strong similarity between the fact that quantum events do not belong to a single Boolean algebra and the fact that observational facts do not belong to a single observer-independent domain. Just as quantum probabilities must be evaluated relative to measurement contexts, observational facts must be expressed relative to perspectives. As in the quantum case, each context (here, each perspective) retains its classical (Boolean) structure; what is new is the structure governing their combination. The new perspectival logic stands to classical logic as the quantum calculus of probability stands to Kolmogorovian probability theory.

**Acknowledgements:** I want to thank Eric Cavalcanti for many useful discussions around these issues.

## Statements and Declarations

**Funding**: None
**Institutional Review Board statement**: not applicable
**Informed Consent Statement**: not applicable
**Data Availability Statement**: not applicable
**Conflicts of Interest:** The author declares no conflict of interest. The opinions expressed in this paper are those of the author and do not necessarily reflect the views of the people mentioned in the acknowledgements.